\documentclass[%
 reprint,
 superscriptaddress,
 amsmath,amssymb,
 aps,
]{revtex4-2}

\usepackage{graphicx}
\usepackage{dcolumn}
\usepackage{bm}
\usepackage[many]{tcolorbox}
\usepackage{lipsum}
\usepackage{amsmath,amssymb,cancel}

\usepackage{shuffle}
\usepackage{graphbox}
\usepackage{environ}
\usepackage{physics}
\usepackage{tabularx}
\usepackage{MnSymbol}
\usetikzlibrary{shapes.geometric}
\usepackage{asymptote}
\newcommand*{\shifttext}[2]{%
  \settowidth{\@tempdima}{#2}%
  \makebox[\@tempdima]{\hspace*{#1}#2}%
}

\newcommand{\abr}[1]{\langle #1\rangle}

\begin{document}

\title{From squared amplitudes to energy correlators}

\author{Song He}
\email{songhe@itp.ac.cn}
\affiliation{CAS Key Laboratory of Theoretical Physics, Institute of Theoretical Physics, Chinese Academy of Sciences, Beijing 100190, China}
\affiliation{School of Fundamental Physics and Mathematical Sciences, Hangzhou Institute for Advanced Study, UCAS \& ICTP-AP, Hangzhou, 310024, China}
\affiliation{Peng Huanwu Center for Fundamental Theory, Hefei, Anhui 230026, P. R. China}
\author{Xuhang Jiang}%
 \email{xhjiang@itp.ac.cn}
 \affiliation{CAS Key Laboratory of Theoretical Physics, Institute of Theoretical Physics, Chinese Academy of Sciences, Beijing 100190, China}
\author{Qinglin Yang}%
 \email{yangqinglin@itp.ac.cn}
 \affiliation{CAS Key Laboratory of Theoretical Physics, Institute of Theoretical Physics, Chinese Academy of Sciences, Beijing 100190, China}
 \affiliation{Max–Planck–Institut f\"ur Physik, Werner–Heisenberg–Institut, D–85748 Garching bei M\"unchen, Germany}
\author{Yao-Qi Zhang}%
 \email{zhangyaoqi@itp.ac.cn}
 \affiliation{CAS Key Laboratory of Theoretical Physics, Institute of Theoretical Physics, Chinese Academy of Sciences, Beijing 100190, China}
\affiliation{%
School of Physical Sciences, University of Chinese Academy of Sciences, No.19A Yuquan Road, Beijing 100049, China
}

\begin{abstract}
The leading order $N$-point energy correlators of maximally supersymmetric Yang-Mills theory in the limit where the $N$ detectors are collinear can be expressed as an integral of the $1\to N$ splitting function, which is given by the $(N{+}3)$-point squared super-amplitudes at tree level. This provides yet another example that the integrand of certain physical observable -- $N$-point energy correlator-- is computed by the canonical form of a positive geometry -- the (tree-level) ``squared amplituhedron". By extracting such squared amplitudes from the $f$-graph construction, we compute the integrand of energy correlators up to $N=11$ and reveal new structures to all $N$; we also show important properties of the integrand such as soft and multi-collinear limits. Finally, we take a first look at integrations by studying possible residues of the integrand: our analysis shows that while this gives prefactors in front of multiple polylogarithm functions of $N=3,4$, the first unknown case of $N=5$ already involves elliptic polylogarithmic functions with
many distinct elliptic curves, and more complicated curves and higher-dimensional varieties appear for $N>5$. 
\end{abstract}

\maketitle


\section{Introduction}
Much of what we have learnt about Quantum Field Theory (QFT) has been gathered by concrete theoretical data and especially explicit computations of observables such as scattering amplitudes and correlation functions. A particularly important class of observables are the $N$-point correlation function of energy flux~\cite{Basham:1977iq, Basham:1978bw,Basham:1978zq, Basham:1979gh,Chen:2020vvp}, which are infrared-finite objects directly measurable in experiments~\cite{Kinoshita:1962ur, Lee:1964is, OPAL:1990reb,ALEPH:1990vew,SLD:1994yoe, Komiske:2022enw, Chen:2022swd, CMS:2023wcp,CMS:2024mlf, Tamis:2023guc}. Recent years have witnessed significant progress on computing such energy correlators in both QCD \cite{Dixon:2018qgp,Luo:2019nig, Dixon:2019uzg, Chen:2019bpb,Yang:2022tgm,Chen:2023zlx} and the cousin ${\cal N}=4$ super-Yang-Mills theory (SYM) \cite{Hofman:2008ar,Belitsky:2013ofa,Henn:2019gkr,Yan:2022cye,Chicherin:2024ifn}: not only have we learnt a great deal about such observables, but various new structures have also been discovered about the ``integrands" and integrals involved in them. In this letter, we will take a concrete step in the computation of the leading order $N$-point energy correlator in ${\cal N}=4$ SYM.

\begin{figure}
   \moveleft 3.9cm  \hbox{
    \includegraphics{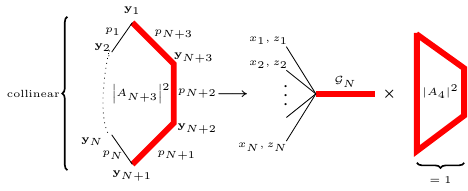}}
   \label{fig:split}
    \caption{The $1\to N$ splitting function from collinear limit of squared amplitudes with $n=N{+}3$ legs.}
\end{figure}

The $N$-point energy correlator in the collinear limit, ${\bf EC}^{(N)}$ depends on complex angles $z_1, z_2, \cdots, z_N$ (with $z_{i,j}:=z_j-z_i$), and it can be written as (permutation sum of) a $(N-1)$-fold integral over {\it energy fractions} $x_1, \cdots, x_N$ of the {\it splitting function} of $1\to N$~\cite{Amati:1978wx,Amati:1978by,Ellis:1978sf,Catani:1998nv}: 
\begin{equation}\label{def_ec}
\begin{aligned}
&{\bf EC}^{(N)}(\{z_i\})=\frac{I_N(z_1, \cdots, z_N)}{|z_{1,2}\cdots z_{N-1,N}|^2}+ {\rm perm}(1,2,\cdots, N)\,,\\
&I_N
:=\int_0^\infty \frac{d^N x}{{\rm GL}(1)}x_{1 2 \cdots N}^{-N}{\cal G}_N (x_1, \cdots, x_N; z_1, \cdots, z_N)
\end{aligned}
\end{equation}
where the $x_i$'s should be regarded as projective coordinates due to the homogeneous weight of ${\cal G}_N$ in each $x_i$, and the measure is defined as $\int d x_1 \cdots x_N \delta(x_{1,\cdots, N}-1)$~\footnote{We denote $x_{i,\cdots, j}:=x_i+ x_{i{+}1}+ \cdots + x_j$; recall that one can instead fix for any subset $I \subset \{1,2,\cdots, n\}$, $\sum_{i\in I} x_i=1$.}; more details can be found in~\cite{Chicherin:2024ifn}. Conventionally the splitting functions at tree level can be obtained from squaring form factors in ${\cal N}=4$ SYM with $N{+}1$ legs~\cite{Chen:2019bpb, Yan:2022cye, Chicherin:2024ifn}, but for our purposes we find it more convenient to consider the squared super-amplitudes with $n=N{+}3$ legs (see Fig.~\ref{fig:split}): in the limit when the first $N$ momenta become collinear, we obtain ${\cal G}_N$ (times $4$-point squared amplitude which is normalized to $1$) as
\begin{equation}\label{split}
{\cal G}_N:=\lim_{1||2\cdots ||N} \frac{|A_n|^2}{|A_{n, {\rm MHV}}|^2}=\lim_{1||2\cdots ||N} \underbrace{\frac12\sum_{k=0}^{n{-}4}\frac{A_{n, k}*A_{n,n{-}4{-}k}}{A_{n,0}*A_{n,n{-}4}}}_{r_n}\,,   
\end{equation}
where we have denoted the squared amplitudes at tree-level (after summing over all helicity sectors) as $|A|_n^2$, and we are really interested in the ratio divided by the overall MHV$*\overline{\rm MHV}$, $r_n$, which turns out to be a dual-conformal-invariant (DCI) function with simple, local poles.  

As shown in Fig.~\ref{fig:split}, we have introduced {\it dual coordinates} for the $n$-gon satisfying ${\bf y}_{i,i{+}1}:={\bf y}_{i{+}1}-{\bf y}_i=p_i$ (for $i=1, \cdots, n$); to take the collinear limit, we parametrize the first $N$ momenta in lightcone coordinates~\cite{Chicherin:2024ifn} using $z_i \to 0$ and $x_i$ as
$p_i=x_i(1,|z_i|^2,z_i)$ (for $i=1,\cdots,N$), while the remaining ones, $p_{i=N{+}1, N{+}2, N{+}3}$ stay generic~\footnote{The collinear limit amounts to $z_i$ becomes small for $i=1,\cdots, N$ or equivalently $z_i \to \infty$ for $i=N{+}1, N{+}2, N{+}3$; additionally one can fix two $z_i$'s to {\it e.g.} $0,1$, thus there are $2(N{-}2)$ real degrees of freedom left.}. The limit can be directly written in terms of the $n(n{-}3)/2=N(N{+}3)/2$ distance squared $(i,j):={\bf y}_{ij}^2$:
\begin{align}\label{collinear}
&(a,b)\to \sum_{a\leq i<j<b} x_i x_j|z_{i,j}|^2\equiv s_{a, 
\cdots, b{-}1},\quad 1\leq a<b{-}1< N{+}1\nonumber\\
&(i, N{+}2)\to x_{i,\cdots, N},\quad i=1,\cdots, N\nonumber\\
&(i{+}1, N{+}3)\to x_{1, \cdots,i},\quad i=1,\cdots,N 
\end{align}   
where $\binom{N}{2}$ of them become quadratic in energy fractions $x_i$ (denoted by $s_{a,...b{-}1}$), while the remaining $2N$ (which involves dual points ${\bf y}_{N{+}2}$ ${\bf y}_{N{+}3}$) are linear (denoted by $x_{i,\cdots j}:=x_i + x_{i{+}1} + \cdots +x_j$).


\section{Three approaches to the squared amplitudes}
\paragraph{Definition} As in \eqref{split}, the DCI ratio $r_n$ is obtained from summing over squared super-amplitudes $A_{n,k}$ for $k=0, 1, \cdots, n{-}4$, which can be computed {\it e.g.} via BCFW recursion directly in momentum twistor space~\cite{Arkani-Hamed:2010zjl}, and the first few examples are
\begin{equation}
\begin{aligned}\label{sqamp_ex}
&r_5=1,\  r_6=1+\frac{A_{6,1}^2}{2 A_{6,2}},\  r_7=1+\frac{A_{7,1}*A_{7,2}}{A_{7,3}},\\
&r_8=1{+}\frac{2 A_{8,1}*A_{8,3}{+}A_{8,2}^2}{2A_{8,4}},\ r_9=1{+}\frac{A_{9,1}*A_{9,4}{+}A_{9,2}*A_{9,3}}{A_{9,5}}.   
\end{aligned}
\end{equation}
It is straightforward to take the $*$ product of $A_{n,k}$ with $A_{n, n{-}4{-}k}$, which is proportional to the $\overline{\rm MHV}$ case~\cite{Dian:2021idl}, and obtain the corresponding ratio (see~\cite{Heslop:2018zut, Dian:2021idl} for such a computation for the $n=6$ example). However, it quickly becomes impractical to cancel spurious poles and reach at a local expression. While in principle we have analytic control of $r_n$ via squaring amplitudes in BCFW form, such a representation usually obscures the surprising simplicity and nice properties of the result,  which is why we propose to compute the squared amplitudes directly (without even computing all the amplitudes). 


\paragraph{The squared amplituhedron} Very nicely, this can be done based on a remarkably simple geometry known as the squared amplituhedron~\cite{Eden:2017fow, Dian:2021idl, Eden:2017fow, He:2024xed}, whose definition is much simpler than the amplituhedra (of various $k$ sectors) themselves~\cite{Arkani-Hamed:2012zlh,Arkani-Hamed:2013jha,Arkani-Hamed:2017tmz}. At tree level, the definition is simply given by the positivity of ``local" poles in {\it bosonized momentum twistor} space, {\it i.e.} $\langle Y i i{+}1 j j{+}1\rangle>0$ except for boundary cases with $j=n$ (for which both signs are included); the sum over $k=0,1, \cdots, n{-}4$ means that we do not need to consider {\it sign flips}~\cite{Arkani-Hamed:2017vfh} and the resulting geometry is much simpler than those for each $k$ sector. The canonical form of this positive geometry, divided by the well-known (bosonized) $\overline{\rm MHV}$ amplitude, $\Omega(Gr_+(n{-}4,n))$, is conjectured to give $r_n$ and it makes properties of the result such as pole structures manifest. This is a beautiful definition of squared amplitudes to all $n$, which will be elaborated elsewhere~\footnote{It is straightforward to compute the canonical form and $r_n$ for {\it e.g.} $n\leq 8$, but the method based on triangulation/ansatz needs to be improved for computing higher-point cases.}. In the following, we will not directly use the geometry but use a method that is more efficient at this stage. 

\paragraph{Null limit of correlator via $f$ graphs}It is well known that $r_n$ can be obtained directly from the $n$-gon null limit of the integrand of $4$-point stress-tensor multiplet correlator at $n{-}4$ loops~\cite{Alday:2010zy,Eden:2010zz,Eden:2010ce,Eden:2011yp,Eden:2011ku,Adamo:2011dq, Heslop:2022xgp}; the latter has been determined up to ten loops via the $f$-graph construction in~\cite{Bourjaily:2016evz} which then allows us to get $r_n$ up to $n=14$: 
\begin{equation}\label{sqamp_f}
r_n=
\xi^{(n)} F^{(n)}|_{\{(1,2)=(2,3)=\cdots=(n,1)=0\}}
\end{equation}
where $F^{(n)}$ denotes the correct combination of $n$-point $f$ graphs~\cite{Bourjaily:2015bpz,Bourjaily:2016evz}; the prefactor $\xi^{(n)}:=\prod_{i=1}^n (i,i{+}1) (i, i{+}2)$ and the lightlike limit is taken with $(i,i{+}1)\to 0$ for $i=1, \cdots, n$. In~\cite{toapp}, we will see that just by requiring even the simplest properties of $r_n$, namely the ``soft" limit (see below), one could push the frontier even further, thus all information about correlator integrand to high loops are secretly contained in this tree-level object! Let us see how this works explicitly for $n\leq 7$, where in each case we have exactly one $f$ graph. The first non-trivial case is for $n=6$~\footnote{For $n=5$, there is a unique $n$-gon lightlike limit which gives $r_5=1$.}, where we find exactly two inequivalent ways of taking the $n$-gon lightlike limit: 
\begin{equation}\label{eq:fig6}
\includegraphics[scale=0.4,align=c]{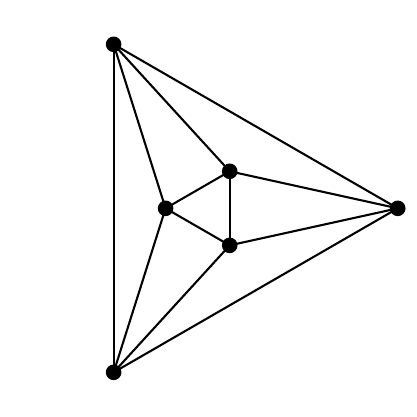}\Longrightarrow\left\{1,
\begin{tikzpicture}[baseline={(0, 0cm )}]
    \foreach \a/\text in {0/1,60/2,120/3,180/4,240/5,300/6} 
    \draw[fill] (\a:1) circle (0.08) node[below] () {\text};
    \draw(0:1)--(180:1);
    \draw(60:1)--(240:1);
    \draw[dashed](60:1)--(180:1);
    \draw[dashed](0:1)--(240:1);
\end{tikzpicture}
\right\}\nonumber
\end{equation}
(with solid/dash line for pole/numerator), and we obtain
\begin{equation}
r_6=1+\frac{(2,4)(1,5)}{(1,4)(2,5)}+ \frac{(2,6)(3,5)}{(3,6)(2,5)}+\frac{(4,6)(1,3)}{(1,4)(3,6)}\,,\nonumber
\end{equation} 
where the two dihedral-inequivalent seeds are $1$, and $\frac{(2,4)(1,5)}{(1,4)(2,5)}$. While for $n=7$ there are three such seeds:
\begin{equation}
\frac{(2,4)(1,5)}{(1,4)(2,5)},\frac{(1,6)^2 (2,4) (3,5)}{(1,4)(1,5)(2,6)(3,6)},\frac{(1,3)(1,6)(2,5)(4,6)}{(1,4)(1,5)(2,6)(3,6)},\nonumber
\end{equation}
\begin{equation}
\includegraphics[scale=0.7,align=c]{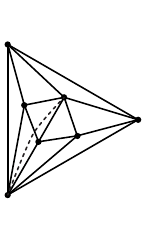}\Rightarrow
\begin{tikzpicture}[scale=0.9,baseline={(0, 0cm )}]
    \foreach \a/\text in {0/1,51.43/2,2*51.43/3,3*51.43/4,4*51.43/5,5*51.43/6,6*51.43/7} 
    \draw[fill] (\a:1) circle (0.08) node[below] () {\text};
    \draw(0:1)--(3*51.43:1);
    \draw(1*51.43:1)--(4*51.43:1);
    \draw[dashed](1*51.43:1)--(3*51.43:1);
    \draw[dashed](0:1)--(4*51.43:1);
\end{tikzpicture},
\begin{tikzpicture}[scale=0.9,baseline={(0, 0cm )}]
    \foreach \a/\text in {0/1,51.43/2,2*51.43/3,3*51.43/4,4*51.43/5,5*51.43/6} 
    \draw[fill] (\a:1) circle (0.08) node[below] () {\text};
    \draw[fill](6*51.43:1.2) circle(0.08)node[below](){7};
    \draw(0:1)--(3*51.43:1);
    \draw(0:1)--(4*51.43:1);
    \draw(1*51.43:1)--(5*51.43:1);
    \draw(2*51.43:1)--(5*51.43:1);
    \draw[dashed](2*51.43:1)--(4*51.43:1);
    \draw[dashed](1*51.43:1)--(3*51.43:1);
    \draw[dashed](0*51.43:1)--(5*51.43:1);
    \draw[dashed](0:1) arc (0:-2*51.43:1) -- (5*51.43:1);
\end{tikzpicture},
\begin{tikzpicture}[scale=0.9,baseline={(0, 0cm )}]
    \foreach \a/\text in {0/1,51.43/2,2*51.43/3,3*51.43/4,4*51.43/5,5*51.43/6,6*51.43/7} 
    \draw[fill] (\a:1) circle (0.08) node[below] () {\text};
    \draw(0:1)--(3*51.43:1);
    \draw(0*51.43:1)--(4*51.43:1);
    \draw(1*51.43:1)--(5*51.43:1);
    \draw(2*51.43:1)--(5*51.43:1);
    \draw[dashed](0*51.43:1)--(2*51.43:1);
    \draw[dashed](0*51.43:1)--(5*51.43:1);
    \draw[dashed](1*51.43:1)--(4*51.43:1);
   \draw[dashed](3*51.43:1)--(5*51.43:1);
\end{tikzpicture}\nonumber
\end{equation}
and each of them generates $7$ distinct terms under the dihedral group, which gives $21$ terms in total. Similarly, we find $22$ dihedral seeds which generate $181$ terms for $n=8$, and we record this and all the seeds up to $n=12$ in an ancillary file. In Table~\ref{tab:staf1} we summarize statistics (number of terms {\it etc.}) for $r_n$ up to $n=13$.
\begin{table}
    \centering
    \begin{tabular}{c|c|c|c|c|c|c|c|c}
    \hline
         n & 6 & 7 & 8 & 9 & 10 & 11 & 12 & 13\\
         \hline
         terms & 4 & 21 & 181 & 2085 & 29016 & 464640 & 9105364& 209639703\\
         \hline
         seeds & 2 &3 & 22 & 134 & 1574 & 21423 & 377307 & 7811985\\
         \hline
         $f$ graphs& 1 & 1 & 3 &7 & 26 & 127 & 1060 & 10525\\
         \hline
    \end{tabular}
    \caption{Some statistics for $r_n$ up to $n=13$: number of terms, dihedral seeds and $f$ graphs with non-zero coefficients.}
    \label{tab:staf1}
\end{table}

\section{New results for EC integrands}
\paragraph{Explicit results and new structures}
It is straightforward to obtain ${\cal G}_N$ from $r_{n=N{+}3}$ by \eqref{collinear}. These integrands for ${\bf EC}^{(N)}$ are known for $N=3,4$~\footnote{See supplementary material for ${\cal G}_{N=4}$ which has $7$ terms with $2$ poles, and $14$ terms with $4$ poles; these expressions agree with those obtained in~\cite{Chicherin:2024ifn} from form factors once we identify the dual points.}, {\it e.g.}
\begin{equation}
{\cal G}_3=1+\frac{x_1 x_3}{x_{1,2} x_{2,3}} +\frac{s_{1,2} x_{1,2,3}}{s_{1,2,3}x_{1,2}} +    \frac{s_{2,3} x_{1,2,3}}{s_{1,2,3}x_{2,3}}\,,
\end{equation}
and now explicit results are available up to $N=11$. 

We also observe interesting structures to all $N$, which can be proven from the geometry or the $f$ graphs. ${\cal G}_N$ has in total $(N{-}1)(N{-}2)/2$ poles that are quadratic in $x_i$'s, $s_{a,a{+}1, \cdots, b{-}1}$ (for $b>a{+}2$) and $2(N{-}2)$ linear ones $x_{1, \cdots, i}$ and $x_{i, \cdots, N}$ for $i=2,\cdots, N{-}1$; individual $x_i$ factors, from the special $s_{a, a{+}1}=x_a x_{a{+}1} |z_{a, a{+}1}|^2$, only appear in the numerators
. Note that in each term we do have {\it incompatible} poles, {\it i.e.} $(i,j)$ and $(k,l)$ which are two intersecting chords in the $n$-gon, but no three poles that are incompatible to each other can appear in the same term, which simply follows from planarity of $f$ graphs; such a property reflects the fact that it is the squared amplitudes, and it puts very strong constraints on possible pole structures of ${\cal G}_N$. 

For any term, the total $x$-weight in the numerator and denominator cancels, and one can show from $f$ graphs that the ``longest" terms contain $2(N{-}2)$ poles, where for $N>3$ we can have up to $2(N{-}3)$ quadratic poles (which we denote as $(2N{-}6) + 2$, $(2N{-}7)+3$ {\it etc.}). For example, for $N=4$, such a term with $2$ quadratic poles and $2$ linear ones ($2+2$ type) reads $\frac{s_{2,3}^2 x_{1,2,3,4}^2}{s_{123} s_{234} x_{1,2,3} x_{2,3,4}}$, while for $N=5$, a term $\frac{s_{12}^2 s_{2345} s_{45} x_5 x_{1,2,3,4,5}}{s_{123} s_{345} s_{1234} s_{12345} x_{3,4,5} x_{2,3,4,5}}$ contains $4$ quadratic poles and $2$ linear ones ($4+2$ type). Some statistics about the pole structures for these ``longest" terms in ${\cal G}_N$ are summarized in Table~\ref{tab:staf2}.


\paragraph{Soft and collinear limits} We comment on some universal behavior of EC integrands. The most basic one follows from the soft limit of the squared amplitudes: as leg $i$ becomes soft or equivalently ${\bf y}_i\to {\bf y}_{i{-}1}$ the $n$-pt squared amplitude reduces to {\it twice} of the $(n{-}1)$-pt one: 
\begin{equation}\label{soft}
\lim_{{\bf y}_i\to {\bf y}_{i{-}1}} r_n=2~r_{n{-}1}(1,\cdots, i{-}1, i{+}1, \cdots, n),
\end{equation}
which will be exploited to constrain $f$ graphs to many points (or four-point correlator integrands to many loops)~\cite{toapp}. Here we remark that by \eqref{soft} when any energy becomes soft  say, $x_N \to 0$, we have ${\cal G}_N \to 2 {\cal G}_{N{-}1}$.  

Another important property is the multi-collinear limit, which essentially follow from the definition of  splitting function:
\begin{equation}
\begin{aligned}
    &\lim_{\substack{z_{1},\cdots, z_{m} \sim \epsilon}} \!\!{\cal G}_{N}=2\,\,\mathcal{G}_{m}(x_{1},\cdots,x_{m};z_{1},\cdots,z_{m})\times \\
    &\hspace{1em}\mathcal{G}_{N-m+1}(x_{1\cdots m},x_{m+1},\cdots,x_{n};z_{m},z_{m+1},\cdots,z_{N})+\mathcal{O}(\epsilon) .
\end{aligned}
\end{equation}
where the leading order in $\epsilon$ for the collinear limit of $1, \cdots, m$ is given by ${\cal G}_m$ times ${\cal G}_{N{-}m{+}1}$ with an additional leg carrying energy fraction $x_{1\ldots m}=x_{1}+\cdots+x_{m}$. Very nicely, such limits of EC integrands immediately leads to collinear limits of ${\bf EC}^{(N)}$, since the phase space integration also splits nicely:
\begin{equation}
\begin{aligned}
    \int\frac{d^{N}x}{GL(1)}x_{12\cdots N}^{-N}= &\int_{0}^{\infty}\frac{dx_{1\cdots m}dx_{m+1}\cdots dx_{N}}{\left(x_{1\cdots m}+x_{m+1}+\cdots+x_{N}\right)^{N}} \\ 
    &\times\int_{0}^{x_{1\cdots m}} dx_{1}\cdots dx_{m-1}\delta(1-x_{1\cdots m}),\nonumber
\end{aligned}
\end{equation}
where we have chosen to fix the redundancy with $\delta(1-x_{1\cdots m})$, and we conclude (at leading order in $\epsilon\to 0$):
\begin{equation}
    \begin{aligned}
        I_{N}& \to 2\int_{0}^{1}d^{m}x \mathcal{G}_{m} \times \int_{0}^{\infty}\frac{dx_{m+1}\cdots dx_{N}}{(1+x_{m+1}+\cdots+x_{N})^{N}}\mathcal{G}_{N-m+1} \\
        &=2I_{m}(x_{1},\cdots,x_{m})\!\times\! \tilde{I}_{N-m+1}(x_{1\cdots m},x_{m+1}\cdots,x_{N})\nonumber 
    \end{aligned}
\end{equation}
where 
$\tilde{I}_{N-m+1}$ indicates that we have deformed $x_{12\cdots N}^{-(N-m+1)}$ to $x_{12\cdots N}^{-N}$ and set $x_{1}=1$ in the definition of ${I}_{N-m+1}$. 
A special case is that when $N-m+1=2$, $\mathcal{G}_{2}=1$ and $\tilde{I}_{2}=\frac{1}{N{-}1}$. 

The full energy correlator $\mathbf{EC}^{(N)}$ is given by a permutation sum (with coefficients like $1/|z_{1,2}\cdots z_{N-1,N}|^2$), at leading order in limit $z_{1},\cdots,z_{m}\to 0$ we only need terms where the permutation factorizes to one that acts on $1,2,\cdots, m$ and one on the remaining labels (and $\{1,\cdots, m\}$ combined). The upshot is
\begin{equation}
    \begin{aligned}
        \lim_{\substack{z_{1},\cdots, z_{m} \sim \epsilon}}\mathbf{EC}^{(N)}=2\,\,\mathbf{EC}^{(m)}\times \widetilde{\mathbf{EC}}^{(N-m+1)} + \mathcal{O}(\epsilon).
    \end{aligned}
\end{equation}
Here $\widetilde{\mathbf{EC}}^{(N-m+1)}$ is defined in a similar way as $\mathbf{EC}^{(N)}$,
\begin{equation}
    \widetilde{\mathbf{EC}}^{(N-m+1)}\!\!\equiv\frac{1}{|z_{m,m+1}\cdots z_{N-1,N}|^2}\tilde{I}_{N-m+1}+\mathrm{perm}(m,\cdots,N).\nonumber
\end{equation}
We note that the definition of $\widetilde{\mathbf{EC}}^{(N-m+1)}$ depends not on $N\!-\!m$ but separately on both $N$ and $m$. When $N=m+1$, $\widetilde{\mathbf{EC}}^{(2)}=\frac{2}{(N{-}1)|z_N|^2}$, and we have
\begin{equation}
    \lim_{\substack{z_{1},\cdots, z_{N{-}1} \sim \epsilon}}\mathbf{EC}^{(N)} =\frac{4}{(N-1)|z_N|^2}\mathbf{EC}^{(N{-}1)}+\mathcal{O}(\epsilon).
\end{equation}
\begin{widetext}
\begin{center}
\begin{table}[h]
    \begin{tabular}{|c|c|c|c|c|c|c|c|c|}
    \hline
         N & 3 & 4 & 5 & 6 & 7 & 8 \\
         \hline
         poles & 1+2 & 3+4 & 6+6 & 10+8 & 15+10 & 21+12\\
         \hline
         terms & 4 & 21 & 181 & 2085 & 29016 & 464640 \\
         \hline
         longest pole st. &1+1 & 2+2, 1+3 &  4+2, 3+3, 2+4 & 6+2, $\cdots$ 
         ,3+5 & 8+2, $\cdots$,4+6 & 10+2, $\cdots$, 5+7  \\
         \hline
         longest terms& 3 &  7 &  36 & 144 & 655 & 2992 \\
         \hline
    \end{tabular}
    \caption{Some statistics for ${\cal G}_N$: the total number of quadratic + linear poles (and the number of terms); possible pole structures and counting for the longest terms.}
    \label{tab:staf2}
\end{table}
\end{center}
\end{widetext}

\section{Towards integrations: residues, elliptic curves and beyond}
Our results for the integrands have already provided numeric access to $\mathbf{EC}^{(N)}$ up to $N=11$, and an efficient way for doing numeric integrations would be highly desirable~\footnote{It is straightforward to perform numerical integrations with high precision {\it e.g.} for $N=5$. We thank Lilin Yang for comments on this. }For general $N$, we expect it to be an arbitrarily complicated task to perform integrations analytically which give transcendental functions of $2(N{-}2)$ independent $z_i$ variables. Nevertheless one can still extract valuable information about the integrated result {\it e.g.} from computing its multi-variable residues, in a way similar to leading singularity/maximal cut analysis for amplitudes or Feynman integrals~\cite{Arkani-Hamed:2010pyv,Dlapa:2021qsl}. We will see that in a way analogous to Feynman integrals, such an analysis not only tells us what kind of transcendental functions we may encounter~\cite{Bourjaily:2019hmc}, but also their physical singularities (poles or branch cuts) and in some cases even possible prefactors and so-called ``symbol letters"~\cite{Goncharov:2010jf} {\it e.g.} via Landau or ``Schubert" analysis~\cite{Landau:1959fi,Arkani-Hamed:2010pyv,Dennen:2016mdk,Yang:2022gko,Fevola:2023fzn,He:2023umf}.

Indeed we have checked that for $N=3,4$, where the integrated results are known to evaluate to MPL functions up to weight $N{-}1$, maximal residues of the integrands account for prefactors of the maximal-weight parts~\cite{Chen:2019bpb, Chicherin:2024ifn}. However, for $N\geq 5$, we will see from a similar analysis that elliptic curves and more complicated varieties start to appear due to the presence of corresponding residues.



We have computed all maximal residues of ${\cal G}_{N=5}$, and in the cases where we could indeed cut $4$ poles (including so-called ``composite" residues), we already encounter solutions to equations from linear all the way up to degree $6$~\footnote{For $N=4$ we encounter cubic roots {\it e.g.} by solving $s_{123}=s_{234}=x_{1234}=0$, which account for all such prefactors of the weight-$3$ part.}, where the $6$-th roots come from conditions of the type $s_{123}{=}s_{234}{=}s_{12345}{=}x_{12345}{=}0$. Even more interestingly, we have encounter cases where we can only cut $3$ poles which lead to elliptic curves! We run into the following {\it four} triplets of conditions: 
\begin{equation}
\{s_{123}=0/s_{1234}=0, s_{234}=0/s_{2345}=0, x_{12345}=0\}\,.    
\end{equation}
For each triplet, one can first solve $x_1$ and $x_5$ from the two quadratic conditions $s_A=s_B=0$, then the linear factor $x_{12345}$ becomes a cubic polynomial of the projective variables $\{x_2,x_3,x_4\}$. A cubic term $x_i^3$ is present if and only if the index $i$ is shared by $A,B$, thus the four cases can be divided into two classes: for $s_{123}=s_{345}=0$, $s_{1234}=s_{345}=0$, or $s_{123}=s_{2345}=0$, $x_{12345}$ is still a quadratic for at least one variable, $x_2$ or $x_4$; while for the last case $s_{1234}=s_{2345}=0$ it is cubic in all three variables. 

For any of the first three cases, one can compute the residues with respect to the variable $x_2$ or $x_4$ and use GL$(1)$ to set $x_3\to1$; this gives a factor $1/\sqrt{P(x_i)}$ where $P(x_i)$ a quartic polynomial in the last variable $x_i$ ($i=2$ or $i=4$), and we cannot take further residues. This gives an elliptic curve $y^2=P(x)$ and it suggests that the integration of such terms should evaluate to elliptic multiple polylogarithm functions (eMPL)~\cite{Broedel:2017kkb,Broedel:2017siw}. In the supplementary material, we will show by direct integration that terms like $\frac{s_{12}s_{45}}{x_{12345}^5 s_{123}s_{345}}$, $\frac{s_{12}s_{34}}{x_{12}x_{12345}^4s_{1234}s_{345}}$ {\it etc.}, which do appear in ${\cal G}_5$, indeed produce eMPL functions with three elliptic curves given by our residue computation.


For the last case, which appears in terms such as $\frac{s_{12345}s_{234}}{x_{12345}^5s_{1234}s_{2345}}$, we cut
\begin{equation}\label{cubicrigidity}
    \{s_{1234}=0, s_{2345}=0, x_{12345}=0\},
\end{equation}
and after solving $s_{1234}=s_{2345}=0$ we have
\begin{equation}
x_{12345}\propto |z_{1,2}z_{2,5}|^2x_2^3+|z_{1,3}z_{3,5}|^2x_3^3+|z_{1,4}z_{4,5}|^2x_4^3+\cdots
\end{equation}
Naively we still have cubic terms in both $x_2$ and $x_4$ after setting $x_3\to1$, but we can do a change of variable $x_4\to a x_2{+}x_4^\prime$ where $a$ satisfies a cubic equation, which cancels $x_2^3$ and one can take another residue since $x_{12345}$ is quadratic in $x_2$. This results in a factor $1/\sqrt{Q(x_4^\prime)}$, with $Q(x_4')$ a quartic polynomial in $x_4'$. Thus we have the fourth elliptic curve, and one can check that direct integration indeed gives eMPL with such a curve. 

We conclude that $I_5$ contains four distinct elliptic curves, but there are a lot more curves in ${\cal G}_5$ after permutation sum. 
For instance, $s_{123}=s_{345}=0$ is invariant under $1\leftrightarrow 2$, $4\leftrightarrow 5$, as well as $\{1\leftrightarrow 5, 2\leftrightarrow 4\}$, thus after $5!$ permutations, the final result contain eMPL with $15$ different elliptic curves; the next two triplets are related up to a reflection, and we have $5!/(2*2){=}30$ curves from them. Finally, $5!/(3!*2)=10$ more elliptic curves are from permutation of the last case. 

We end this section with comments for general $N$. One can show that terms with pole structures 
\[\{s_{1,\cdots,j},s_{i,\cdots,N},x_{1\cdots N}\},\ i>j\]
always appear in ${\cal G}_N$. Again for cases that at least one index $a$ is not shared in the triple, on the support of $s_{1,\cdots,j}{=}s_{i,\cdots,N}{=}0$, $x_{1\cdots N}$ is quadratic in $x_a$, thus after computing the triple residues we have a factor $1/\sqrt{P(x_2,\cdots,\hat{x}_a,\cdots,x_{N{-}1})}$ and no further residue can be taken. This corresponds to a surface of dimension $N-4$ and it is expected that integration of such term has ``rigidity" $N{-}4$ (with rigidity-$1$ case corresponds to a curve)~\cite{Bourjaily:2018ycu}. For the maximally overlapped case with triplet $\{s_{1,\cdots, N{-}1},s_{2,\cdots, N},x_{1\cdots N}\}$, after solving $s_{1,\cdots, N{-}1}=s_{2,\cdots, N}=0$, we have a polynomial cubic in all variables \[x_{1\cdots N}\propto\sum_{i=2}^{N{-}1}|z_{1,i}z_{i,N}|^2 x_i^3+\cdots\]
and a similar shift $x_{N{-}1}\to a x_2{+} x_{N{-}1}^\prime$ also makes $x_{1\cdots N}$ quadratic in $x_2$; we still end up with a similar $(N{-}4)$-dimensional surface $\sqrt{Q(x_3,\cdots,x_{N{-}1}^\prime)}$, which suggests the same rigidity. We do not know if such surfaces for $N>5$ correspond to Calabi-Yau geometry~\cite{Bourjaily:2018ycu,Bourjaily:2018yfy,Vergu:2020uur,Cao:2023tpx} or not, and we leave the investigations to future work. Note that there are terms that contain the triplet but also additional linear poles: for them we can perform more cuts to reduce the rigidity, and we will end up with lower-dimensional surfaces; in supplementary material we will study such cases for $N=6$ where we obtain hyper-elliptic curves with genus two.

\section{Conclusions and outlook}
We have proposed that the integrands for energy correlators with $N$ collinear detectors at leading order in $\mathcal{N}=4$ SYM theory can be obtained from the squared amplituhedron geometry or equivalently the null limit of stress-tensor multiple correlator, which via the $f$ graphs has provided data up to $N=11$ and allowed us to uncover new structures to all $N$. We have studied their general pole structures, and universal behavior under soft and multi-collinear limits. We have also taken the first step in integrating them: by residue analysis we have seen that the energy correlator with $N\geq 5$ involves increasingly complicated transcendental functions beyond MPL. 

Our preliminary studies have opened up several new avenues for investigations. Given our path from the squared amplituhedron geometry to integrands and finally to the energy correlators, it is intriguing to ask how much we could learn about integrands and residues to all $N$ directly from the geometry. Since the soft limit and its generalizations put extremely strong constraints on squared amplitudes (and loop integrands of correlators) when combined with $f$ graphs, we hope to determine $r_n$ and ${\cal G}_N$ to even higher points. We have only considered EC in collinear limits, and it would be interesting to explore possible underlying geometry for integrands of EC with generic angles, or the squared form factors. For integrations, it would be highly desirable to apply various methods for direct integrations~\cite{Bourjaily:2018aeq,He:2020lcu, Arkani-Hamed:2017ahv} or integration by parts~\cite{Caron-Huot:2014lda,Henn:2022vqp} to this rather difficult problem; relatedly our residue analysis can be regarded as the first step in bootstrapping these observables based on prefactors and symbol letters~\cite{Caron-Huot:2016owq,Henn:2018cdp,Caron-Huot:2020bkp,Dixon:2020bbt,He:2021non,He:2021eec}. Given the complexities for $N\geq 5$, we would like to understand the geometry of higher-rigidity surfaces and the associated singularity structures~\cite{Bourjaily:2019hmc,Vergu:2020uur}, and even develop some notion of ``symbology" similar to elliptic case~\cite{Kristensson:2021ani,Wilhelm:2022wow,Morales:2022csr, He:2023qld,Spiering:2024sea}; it might also be fruitful to consider alternative representation for $\mathbf{EC}^{(N)}$, such as Mellin representation (like those for Witten diagrams in AdS space~\cite{Mack:2009mi,Fitzpatrick:2011ia}) to manifest certain analytic properties of the integrated result. Finally, it would be fascinating to shed new lights on all these beautiful interconnections between energy correlators, stress-tensor multiplet correlators, and amplitudes/form factors in ${\cal N}=4$ SYM.

\begin{acknowledgments}
We thank Jiahao Liu, Rourou Ma, Kai Yan, Lilin Yang, Yang Zhang and Huaxing Zhu for inspiring discussions or comments on the draft, and Canxin Shi, Yichao Tang for collaborations on related projects. The work of S.H. has been supported by the National Natural Science Foundation of China under Grant No. 12225510, 11935013, 12047503, 12247103, and by the New Cornerstone Science Foundation through the XPLORER PRIZE.
\end{acknowledgments}

\bibliographystyle{apsrev4-1}
\bibliography{bib}

\newpage

\widetext
\begin{center}
\textbf{\Large Supplementary Material}
\end{center}
\section*{Explicit results for EC integrands}
Here we briefly summarize our results for the squared amplitudes, $r_n$ and the corresponding splitting function/integrand for energy correlator ${\cal G}_{N=n{-}3}$. From the $f$-graph construction, one can extract $r_n$ up to $n=14$ thus ${\cal G}_N$ up to $N=11$. The size of the explicit results gets large rather quickly: as we have mentioned in the main text, there are over $2\times 10^8$ terms in $r_{13}$ or ${\cal G}_{10}$; Due to size limits for arXiv submission, we have only included (seeds for) $r_{n\leq 12}$ and ${\cal G}_{N\leq 7}$ (higher-point results are available upon request). 

For $r_n$ it is convenient to record all dihedral-inequivalent seeds, and we can also write such DCI expressions in terms of {\it cross-ratios}. A basis of {\it multiplicatively independent} cross-ratios (which still satisfy Gram-determinant constraints) are $u_{i,j}:=\frac{(i, j{+}1)(i{+}1, j)}{(i,j) (i{+}1, j{+}1)}$ for $|j-i|>2$ (the total number is $n(n{-}5)/2$
). In this way, the $n=6$ squared-amplitude reads
\begin{equation}
1+\frac{(2,4)(1,5)}{(1,4)(2,5)}+ \frac{(2,6)(3,5)}{(3,6)(2,5)}+\frac{(4,6)(1,3)}{(1,4)(3,6)}=1+ u_{1,4} + \text{dihedral no repeat}
\end{equation}
where we have written it in terms of two {\it seeds}, $1$ and $u_{1,4}$ where the dihedral transformation of $n=6$ give $1$
and $3$ terms without repeat, respectively.  For $n=7$ we have $21$ terms which can be written in terms of $3$ seeds:
\begin{equation}
\frac{(2,4)(1,5)}{(1,4)(2,5)}+ \frac{(1,6)^2 (2,4) (3,5)}{(1,4)(1,5)(2,6)(3,6)}+\frac{(1,3)(1,6)(2,5)(4,6)}{(1,4)(1,5)(2,6)(3,6)} +\text{d.n.r}=u_{1,4} + u_{1,4} u_{1,5}^2 u_{2,5} + u_{1,4} u_{1,5} u_{3,6}+\text{d.n.r}
\end{equation}
As we have mentioned, by plugging \eqref{collinear} we obtain the splitting function for $N=3,4$, {\it e.g.}
\begin{align*}
&{\cal G}_4=\frac{x_1 x_{3,4}}{x_{1,2} x_{2,3,4}}+\frac{x_4 x_{1,2}}{x_{3,4} x_{1,2,3}}+\frac{s_{1,2} x_{1,2,3}}{s_{1,2,3} x_{1,2}}+\frac{s_{3,4} x_{2,3,4}}{s_{2,3,4} x_{3,4}}+\frac{s_{1,2,3} x_{1,2,3,4}}{s_{1,2,3,4} x_{1,2,3}}+\frac{s_{2,3,4} x_{1,2,3,4}}{s_{1,2,3,4} x_{2,3,4}}+\frac{s_{2,3} s_{1,2,3,4}}{s_{1,2,3} s_{2,3,4}}\\
&+\frac{s_{1,2} s_{2,3} x_{1,2,3,4}^2}{s_{2,3,4} s_{1,2,3,4} x_{1,2} x_{1,2,3}}+\frac{s_{2,3} s_{3,4} x_{1,2,3,4}^2}{s_{1,2,3} s_{1,2,3,4} x_{3,4} x_{2,3,4}}+\frac{x_1 s_{3,4}^2 x_{1,2,3,4}}{s_{2,3,4} s_{1,2,3,4} x_{1,2} x_{3,4}}+\frac{x_4 x_1^2 s_{3,4}}{s_{2,3,4} x_{1,2} x_{1,2,3} x_{2,3,4}}+\frac{x_4^2 x_1 s_{1,2}}{s_{1,2,3} x_{3,4} x_{1,2,3} x_{2,3,4}}\\
&+\frac{x_4 s_{1,2}^2 x_{1,2,3,4}}{s_{1,2,3} s_{1,2,3,4} x_{1,2} x_{3,4}}+\frac{s_{2,3}^2 x_{1,2,3,4}^2}{s_{1,2,3} s_{2,3,4} x_{1,2,3} x_{2,3,4}}+\frac{x_4 x_1 s_{2,3} s_{1,2,3,4}}{s_{1,2,3} s_{2,3,4} x_{1,2,3} x_{2,3,4}}+\frac{x_1 s_{2,3} x_{3,4} x_{1,2,3,4}}{s_{2,3,4} x_{1,2} x_{1,2,3} x_{2,3,4}}+\frac{x_4 s_{2,3} x_{1,2} x_{1,2,3,4}}{s_{1,2,3} x_{3,4} x_{1,2,3} x_{2,3,4}}\\
&+\frac{x_4 s_{1,2} s_{2,3,4} x_{1,2,3,4}}{s_{1,2,3} s_{1,2,3,4} x_{3,4} x_{2,3,4}}+\frac{x_1 s_{3,4} s_{1,2,3} x_{1,2,3,4}}{s_{2,3,4} s_{1,2,3,4} x_{1,2} x_{1,2,3}}+\frac{s_{1,2} s_{3,4} x_{1,2,3} x_{1,2,3,4}}{s_{1,2,3} s_{1,2,3,4} x_{1,2} x_{3,4}}+\frac{s_{1,2} s_{3,4} x_{2,3,4} x_{1,2,3,4}}{s_{2,3,4} s_{1,2,3,4} x_{1,2} x_{3,4}}
\end{align*}

Similarly for $n=8$, we have $22$ seeds which gives a total of $181$ terms (note that their coefficients can be $\pm 1$)
\begin{equation}
\begin{aligned}
    &{-}1{+}\frac{(2,7) (3,6)}{(2,6) (3,7)}{-}\frac{(1,7) (4,6)}{(1,6) (4,7)}{-}\frac{(1,7) (2,4) (5,7)}{(1,5) (2,7) (4,7)}{+}\frac{(1,7) (2,6)^2 (3,5)}{(1,6) (2,5) (2,7) (3,6)}{+}\frac{(2,7)^2 (3,5) (4,6)}{(2,5) (2,6) (3,7) (4,7)}{+}\frac{(1,3) (2,5) (6,8)^2}{(1,6) (2,6) (3,8) (5,8)}\\
    &{+}\frac{(2,6) (2,8) (3,5) (3,7)}{(2,5) (2,7) (3,6) (3,8)}{-}\frac{(1,3) (2,8) (4,6) (5,7)}{(1,4) (2,7) (3,6) (5,8)}{+}\frac{(1,7) (2,5) (4,8) (5,7)}{(1,5) (2,7) (4,7) (5,8)}{+}\frac{(1,3) (2,7) (3,6) (6,8)}{(1,6) (2,6) (3,7) (3,8)}{+}\frac{(1,3) (2,5) (4,6) (6,8)^2}{(1,6) (2,6) (3,6) (4,8) (5,8)}\\
    &{+}\frac{(1,3) (2,4) (2,8) (5,7)^3}{(1,5) (2,5) (2,7) (3,7) (4,7) (5,8)}
    {+}\frac{(1,3) (2,4)^2 (5,7) (6,8)^2}{(1,6) (2,5) (2,6) (3,8) (4,7) (4,8)}
    {+}\frac{(1,3) (1,7) (2,6)^2 (3,5) (5,7)}{(1,5) (1,6) (2,5) (2,7) (3,6) (3,7)}\\
    &{+}\frac{(1,3) (2,6) (2,8) (3,5) (5,7)^2}{(1,5) (2,5) (2,7) (3,6) (3,7) (5,8)}+\frac{(1,3) (1,7) (2,5)^2 (4,6) (6,8)}{(1,4) (1,5) (2,6) (2,7) (3,6) (5,8)}+\frac{(1,7) (2,7) (2,8) (3,5) (3,6) (4,6)}{(1,6) (2,5) (2,6) (3,7) (3,8) (4,7)}\\
    &{+}\frac{(1,3) (1,7) (2,6) (3,5) (4,8) (5,7)}{(1,5) (1,6) (2,5) (3,7) (3,8) (4,7)}
    -\frac{(1,3) (1,7) (2,8) (3,5) (4,6) (5,7)}{(1,4) (1,5) (2,7) (3,6) (3,7) (5,8)}+\frac{(1,3) (1,7) (2,5) (3,8) (4,6) (5,7)}{(1,4) (1,5) (2,7) (3,6) (3,7) (5,8)}\\
    &{+}\frac{(1,3) (2,4) (2,7) (3,6) (5,7) (6,8)}{(1,6) (2,5) (2,6) (3,7) (3,8) (4,7)} +\text{d.n.r}\\
    &{=}-1+u_{2,6}+u_{1,6} u_{2,5}+u_{2,5} u_{2,7}-u_{1,6} u_{2,6} u_{3,6}+u_{2,5} u_{2,6}^2 u_{3,6}+u_{1,6} u_{2,5} u_{2,6}^3 u_{2,7} u_{3,6}+u_{1,5} u_{1,6} u_{4,7}+u_{1,5} u_{1,6} u_{3,6} u_{3,7}^2 u_{3,8} u_{4,7}\\
    &{-}u_{2,7} u_{3,6} u_{3,7}^2 u_{3,8} u_{4,7}-u_{1,5} u_{1,6} u_{2,5} u_{2,6} u_{2,7} u_{3,6} u_{3,7}^2 u_{3,8} u_{4,7}-u_{1,4} u_{1,5} u_{1,6} u_{4,7} u_{4,8}+u_{1,6} u_{2,5} u_{2,6} u_{3,8} u_{4,7} u_{4,8}\\
    &{+}u_{1,6} u_{2,5} u_{3,7} u_{3,8} u_{4,7} u_{4,8}+u_{2,5} u_{2,7} u_{3,7}^2 u_{3,8} u_{4,7}^2 u_{4,8}+u_{1,4} u_{2,7} u_{3,7}^2 u_{3,8} u_{4,7}^3 u_{4,8}^2+u_{1,5}^2 u_{1,6} u_{3,6} u_{3,7} u_{3,8} u_{5,8}\\
    &{+}u_{2,6} u_{3,8} u_{4,8} u_{5,8}+u_{1,4} u_{2,6} u_{3,8} u_{4,7} u_{4,8}^2 u_{5,8}+u_{1,5} u_{3,8} u_{4,8} u_{5,8}^2+u_{1,5} u_{3,6} u_{3,7} u_{3,8} u_{4,8} u_{5,8}^2+u_{1,4}^2 u_{1,5} u_{3,8} u_{4,7} u_{4,8}^3 u_{5,8}^2+\text{d.n.r}
    \end{aligned}
\end{equation}

There are many interesting patterns for $r_n$, {\it e.g.} for even $n$ it has a constant term (which goes like $-1, 2, -5, 14$ for $n=8,10, 12, 14$) but such constants cannot exist for odd $n$. We also observe that up to $n=11$, $r_n$ can always be written as the polynomial of $u_{i,j}$. However, starting $n=12$, it contains terms with $u_{i,j}$ in the denominator. All seeds in $u$ variables up to $n=12$ are given in the ancillary file. 

\section*{Details on direct integrations and residues of EC integrands}
\paragraph{Direct integration for eMPL terms}
Let us study how elliptic integrals appear in direct integration for terms in ${\cal G}_5$. We select the following single term 
\begin{equation}\label{elliptic15pt}
    \frac{s_{12}s_{45}}{x_{12345}^5 s_{123}s_{345}}
\end{equation}
of the $N=5$ EC integrand as an example, which contains the first triple we discussed in the main text. Since direct integration for this term is quite lengthy, we firstly decompose this integral through integration by part onto proper master integrals, of which direct integration can be performed more straightforwardly. The number of master integrals and corresponding decomposition can be determined by the method of intersection theory~\cite{Mastrolia:2018uzb,Frellesvig:2019uqt,Frellesvig:2020qot,Weinzierl:2020xyy,Caron-Huot:2021xqj}. Note that here we only need to check whether this basis is independent by calculating their intersection number matrix. The term belongs to the integral family 
\begin{equation}
 \int\frac{{\rm d}^5x_i}{GL(1)}\frac{s_{12}^{-a_4}s_{23}^{-a_5}s_{34}^{-a_6}s_{45}^{-a_7}}{x_{12345}^{a_1} s_{123}^{a_2}s_{345}^{a_3}}   
\end{equation}
spanned by $4$ independent master integrals as
\begin{align}
&A_1{=}\int\frac{{\rm d}^5x_i}{GL(1)}\frac{1}{x_{12345}s_{123}s_{345}}, A_2{=}\int\frac{{\rm d}^5x_i}{GL(1)}\frac{x_1}{x_{12345}^2s_{123}s_{345}},\nonumber\\
&A_3{=}\int\frac{{\rm d}^5x_i}{GL(1)}\frac{1}{x_{12345}s_{123}^2},\ A_4{=}\int\frac{{\rm d}^5x_i}{GL(1)}\frac{1}{x_{12345}s_{345}^2}
\end{align}
$A_3$ and $A_4$ are MPL functions. For $A_1$, integrating $x_1$, $x_2$ and $x_5$, and setting $x_3\to1$, we finally arrive at a one-fold integration for a uniform weight-3 pure MPL function $\mathcal{I}_1^{(3)}(x_4)$ as
\[A_1=\int_0^\infty\frac{{\rm d}x_4}{\sqrt{P(x_4)}}\mathcal{I}_1^{(3)}(x_4)\]
with $P(x_4)$ being the quartic polynomial mentioned in the main text:
\begin{align}
    P(x_4)&=\left(\langle 35\rangle (\langle 12\rangle{+}\langle 13\rangle{-}\langle 23\rangle){+}(\langle 12\rangle\langle 35\rangle {+}\langle 12\rangle\langle 45\rangle{-}\langle 12\rangle\langle 34\rangle  {+}\langle 13\rangle\langle 45\rangle {-}\langle 23\rangle\langle 45\rangle) x_4{+}\langle 12\rangle\langle 45\rangle x_4^2\right)^2\nonumber\\   &-4\langle12\rangle\langle13\rangle(\langle35\rangle+\langle45\rangle x_4)(\langle35\rangle+(\langle35\rangle+\langle45\rangle-\langle34\rangle)x_4 +\langle 45\rangle x_4^2 )
\end{align}
where we introduce the shorthand notation $\langle ij\rangle:=|z_{i,j}|^2$. According to the definition in \cite{Broedel:2017kkb}, $A_1$ yields an eMPL $E_4$ function. Similar functions appear in $A_2$ after integration:
\[  A_2=\int_0^\infty\frac{{\rm d}x_4}{\sqrt{P(x_4)}}\mathcal{I}_2(x_4)\]
where $I_2(x_4)$ is an MPL function up to weight $3$. So we can conclude that \eqref{elliptic15pt} is also eMPL after performing integration by part. Since $A_{1}$ and $A_{2}$ form an elliptic system, we expect them to be connected by partial differentiation. In fact, we can show through integration by part that
\begin{equation}
    A_{2}=A_{1}+\abr{13}\partial_{\abr{13}}A_{1}.
\end{equation}

Note that for those terms containing not only the triple but other factors as poles, cutting extra poles provides further relations between the variables, which leads to a lower rigidity surface after solving the leading singularity. For instance, at $N=5$ we have term \begin{equation}
\frac{s_{12}s_{45}x_{123}x_{345}}{x_{12345}^5s_{1234}s_{2345}s_{123}s_{345}}
\end{equation}
On the support of $s_{123}{=}s_{345}{=}0$ as well as $s_{1234}{=}0$ (or $s_{2345}{=}0$), $x_{12345}{=}0$ turns out to be a quadratic equation, which yields MPL residue finally. However, it should be expected that its integrated result still contains eMPL, since all the resultants from the poles should appear when we perform the integration and partial fraction. 
Similar discussion applies to other two triples: we have terms {\it e.g.} $\frac{s_{12}s_{34}}{x_{12}x_{12345}^4s_{1234}s_{345}}$ in the $N=5$ integrand, and eMPL functions appear after integration as well.

\paragraph{More details on residues and curves}
Next we comment on various residues and (hyper-)elliptic curves for EC integrands. We have seen that not only higher-rigidity algebraic surfaces appear for higher-point energy correlators, but also more involved residues even in the MPL case and {\it e.g.} hyper-elliptic curves~\cite{Huang:2013kh,Hauenstein:2014mda,Marzucca:2023gto} emerge. For the former, we take an example of $N=5$, and consider cuts for the following term appearing in the integrand,
\begin{equation}
    \frac{x_1x_5s_{34}^{2}s_{12345}}{x_{12345}^{5}x_{1234}x_{345}s_{234}s_{345}s_{1234}} \to \frac{s_{34}^{2}}{x_{12345}x_{1234}x_{345}s_{234}s_{345}s_{1234}}.
\end{equation}
The RHS is related to LHS by integration-by-part relations. We only perform an analysis for the RHS.
Multiple cuts exist for above integrand with such denominators. If we cut $x_{345},x_{12345},s_{345}$ and solve for variable $x_1,x_3,x_5$, then after fixing $x_2=1$, a polynomial $P(x_4)$ will appear in the denominator,
\begin{equation}
\begin{aligned}
    P(x_4)=&-2\abr{12}\abr{35}+\Big[2\abr{35}(\abr{24}-\abr{14})+(\abr{13}-\abr{23})\Big(\abr{35}+\abr{45}-\abr{34}-\sqrt{\lambda(\abr{34},\abr{35},\abr{45})}\Big)\Big]x_{4} \\
    & -\abr{34}\Big(\abr{35}+\abr{45}-\abr{34}-\sqrt{\lambda(\abr{34},\abr{35},\abr{45})}\Big)x_4^2,
\end{aligned}
\end{equation}
where $\lambda(\abr{34},\abr{35},\abr{45})=(\abr{35}+\abr{45}-\abr{34})^2-4\abr{35}\abr{45}$. So after solving $x_4$, leading singularities will involve square root of $|z_{i,j}|^2$'s under square root which can be viewed as solutions for some quartic equation. More complicated leading singularities can appear when we study the cut of $x_{12345},s_{234},s_{345},s_{12345}$ which is performed to the following term in the integrand,
\begin{equation}
    \frac{s_{34}s_{1234}s_{2345}}{x_{12345}^{4}x_{1234}s_{234}s_{345}s_{12345}}.
\end{equation}
After solving $x_1,x_3,x_4$ and fixing $x_2=1$, the last integration will involve a polynomial of degree 6 of $x_5$ in the denominator which is too complicated to present here. It indicates that the leading singularities will involve a 6-th root of kinematic variables. 

For the latter, namely the appearance of hyper-elliptic curves, we take an example for $N=6$, and solve the cut for the following term which appear in the integrated:
\begin{equation}
    -\frac{s_{12}s_{45}^2}{x_{123456}^{5}x_{12345}s_{123}s_{345}s_{456}} \to \frac{x_{5}^2}{x_{123456}x_{12345}s_{123}s_{345}s_{456}}.
\end{equation}
The RHS is related to the LHS by integration-by-part relations and we only perform an analysis for the RHS.
After cutting $x_{123456},s_{123},s_{345},s_{456}$ and solving for $x_{1},x_{2},x_{3},x_{6}$, we will arrive at a geometric genus-2 hyperelliptic curve for the integrand of the remaining variable $x_{4}$ (with $x_{5}$ fixed to 1 by the delta function) which can be written as a combination of $\mathrm{d}x_{4}/\sqrt{H(x_{4})}$ and $\mathrm{d}x_{4}/x_4/\sqrt{H(x_{4})}$ where $H(x_{4})$ is a polynomial of degree 6,
\begin{equation}
\begin{aligned}
   H(x_4)=& \Big[\abr{12}\abr{35}\abr{56}+\Big((\abr{
    34
    }-\abr{45})\abr{12}\abr{56}+(\abr{13}-\abr{23})\abr{45}\abr{56}+\abr{12}\abr{35}(\abr{46}+\abr{56}-\abr{45})\Big)x_{4}+\\ 
    &\Big((\abr{
    34
    }-\abr{45})\abr{12}\abr{46}+(\abr{13}-\abr{23})\abr{45}\abr{46}+\abr{12}\abr{34}(\abr{46}+\abr{56}-\abr{45})\Big)x_{4}^2+\abr{12}\abr{34}\abr{46}x_{4}^3\Big]^2+\\
    &4\abr{12}\abr{23}\abr{45}x_4\Big(\abr{46}x_4+\abr{56}x_5\Big)\Big[\abr{35}\abr{56}+\Big((\abr{34}-\abr{45})\abr{56}+\abr{35}(\abr{46}+\abr{56}-\abr{45})\Big)x_4+\\
    &\hspace{18em}\Big((\abr{35}-\abr{45})\abr{46}+\abr{34}(\abr{46}+\abr{56}-\abr{45})\Big)x_4^2+\abr{34}\abr{46}x_4^3\Big].
\end{aligned}
\end{equation}
Similar curve appear when we cut $x_{123456},s_{345},s_{456},s_{12345}$ or $x_{123456},s_{123},s_{234},s_{23456}$ which appears in other terms of the $N=6$ integrand. 
\end{document}